# Countermeasure against tailored bright illumination attack for DPS-QKD

Toshimori Honjo,[1,*] Mikio Fujiwara,[2] Kaoru Shimizu,[3] Kiyoshi Tamaki,[3] Shigehito Miki,[2] Taro Yamashita,[2] Hirotaka Terai,[2] Zhen Wang,[2] and Masahide Sasaki[2]

[1] *NTT Secure Platform Laboratories, NTT Corporation, 3-9-11 Midori-cho, Musashino, Tokyo 180-8585, Japan*
[2] *National Institute of Information and Communication Technology, 4-2-1 Nukui-kitamachi, Koganei, Tokyo 184-8795, Japan*
[3] *NTT Basic Research Laboratories, NTT Corporation, 3-11 Morinosato Wakamiya, Atsugi, Kanagawa 180-8585, Japan*
[*]*honjo.toshimori@lab.ntt.co.jp*

**Abstract:** We propose a countermeasure against the so-called tailored bright illumination attack for Differential-phase-shift QKD (DPS-QKD). By monitoring a rate of coincidence detection at a pair of superconducting nanowire single-photon detectors (SSPDs) which is connected at each of the output ports of Bob's Mach-Zehnder interferometer, Alice and Bob can detect and defeat this kind of attack. We also experimentally confirmed the feasibility of this countermeasure using our 1GHz-clocked DPS-QKD system. In the emulation of the attack, we achieved much lower power of the bright illumination light compared with the original demonstration by using a pulse stream instead of broad pulses.

## 1. Introduction

Quantum key distribution (QKD) is in principle capable of establishing a provably secure key between two remote parties, that is, the probability that an eavesdropper, who has unbounded ability, gets complete information on the key can be set as small as desired [1]. To realize this, several QKD protocols have been proposed such as BB84[2] and E91[3]. One should, however, note that the security proof is based on certain assumptions on its physical implementation, such as the eavesdropper should have no access to Alice's transmitter and Bob's receiver, and their devices should operate as required by the protocol [4].

In practice, however, it is often the case where these assumptions cannot be met. Actually, photon detectors in the receiver can be readily controlled by sending tailored optical pulses through the quantum channel, which was not considered to take place in the security proofs. The attack using this kind of discrepancy between a theoretical model and its physical implementation is usually called side channel attacks. It is a hot issue to identify quantum side channels and prescribe countermeasure for them.

Among various side channel attacks, those related to photon detectors are serious, because they are usually common to the most of QKD schemes. A representative one is blinding attack with bright light, as shown to be effective for semiconductor avalanche photodiode in the two commercial QKD systems [5], and the implementations of countermeasures were reported [6,7].

Recently, L. Lydersen et al. proposed a way to hack a superconducting nanowire single-photon detector (SSPD) with the so-called tailored bright illumination [8,9]. Also they proposed how to perform intercept and resend attack for SSPD employed in Differential-phase-shift (DPS-QKD) system. In this paper, we propose a countermeasure against the tailored bright illumination attack for DPS-QKD. We also experimentally show the feasibility of our countermeasure.

## 2. Differential-phase-shift QKD (DPS-QKD)

First, we briefly explain DPS-QKD scheme [10], whose typical setup and protocol are shown in Fig. 1. Alice creates a coherent pulse train by a coherent light source and an intensity modulator, which is randomly phase-modulated for each pulse with $\{0, \pi\}$, attenuated to be less than one (e.g., 0.2) photon per pulse on average, and then sent to Bob. Bob measures the phase difference between two sequential pulses using a 1-bit delay Mach-Zehnder interferometer and single photon detectors, and records the photon arrival time and which detector clicked. After transmission of the optical pulse train, Bob tells Alice the time instances at which a photon was counted. From this time information and her modulation data,

Alice knows which detector clicked at Bob's site. Under an agreement that a click by detector 1 is assigned to a bit value "0" and a click by detector 2 represents "1", Alice and Bob obtain bit strings (sifted key). After the error correction and privacy amplification, Alice and Bob can share the final key. After the first demonstration of DPS-QKD using planar light wave circuit Mach-Zehnder interferometer (PLC-MZI) [11], we performed several long distance transmission experiments including 200-km transmission with SSPDs [12,13]. Recently, S. Wang et al. successfully demonstrated 260-km transmission with fiber-based interferometer and SSPDs [14]. In a field demonstration in Tokyo QKD Network, a secure key rate at 2kbps over a 90-km installed fiber was demonstrated [15]. As for the security proof, though the security against the general individual attack was proved [16], the unconditional security of the original DPS-QKD protocol has not been proved yet. However, K. Tamaki et al. proved the unconditional security of DPS-QKD protocol with block-wise phase randomization [17].

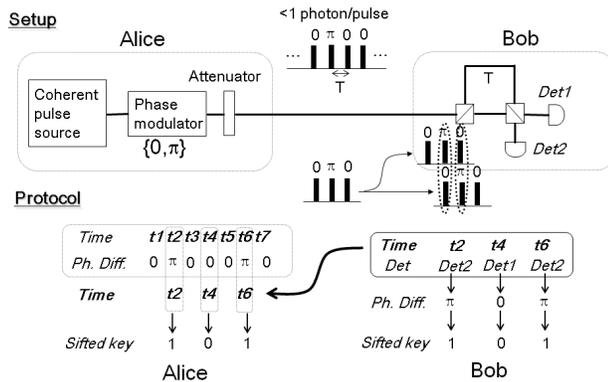

Fig. 1. Setup and protocol of DSP-QKD.

## 3. Tailored Bright Illumination Attack and its Countermeasure

L. Lydersen et al. proposed a way to control the dead time of SSPD by injecting a bright pulse into the detectors [8,9]. As shown in Fig. 2, during the illumination of a sufficiently bright optical pulse, the electrical signal of the output of SSPD stays above the comparator threshold, which is regarded as no detection of photons. Note that SSPD clicks only when the rising edge of the output voltage across the comparator threshold. Furthermore, Eve can make fake counts at any designated time as follows. First, she starts to illuminate the SSPD with a long bright pulse to change the state of the SSPD into the blinding state. Slightly before the time when she wants to make a fake click, she shuts down the illumination. After the SSPD becomes clickable state, she restarts the bright illumination again. At the rising edge of this bright illumination pulse, SSPD outputs a click signal as she desired.

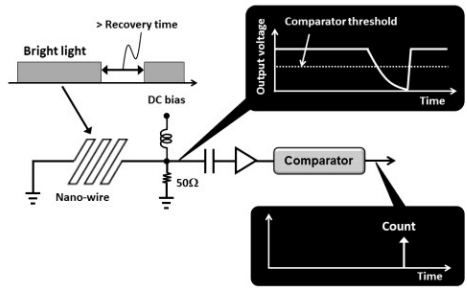

Fig. 2. Tailored bright illumination attack.

By using the above idea, L. Lydersen et al. also proposed the attack for DPS-QKD [8,9]. Figure 3 shows the diagram of their attack. To control the SSPDs behind the Mach-Zehnder interferometer, they apply an appropriate phase modulation to bright pulses. Eve first injects the bright pulse into both detectors by applying a phase of {0, π/2} such that power splits equally to the two detectors, and both detectors get blind ((a) in Fig. 3). Then, Eve inputs a sequence of {0,0} (or {0, π}) phase-modulated short pulses in such a way that one of two SSPDs becomes clickable and the other stays blind ((b) in Fig. 3). After the voltage of the appropriate SSPD falls below the comparator threshold (after the dead time), Eve restarts the bright {0, π/2} phase-modulated pulse to make the SSPD click and change it into blinding state again ((c) in Fig. 3). Eve can perform the so-called intercept and resend attack using this detector control method, and she can perform a perfect eavesdropping.

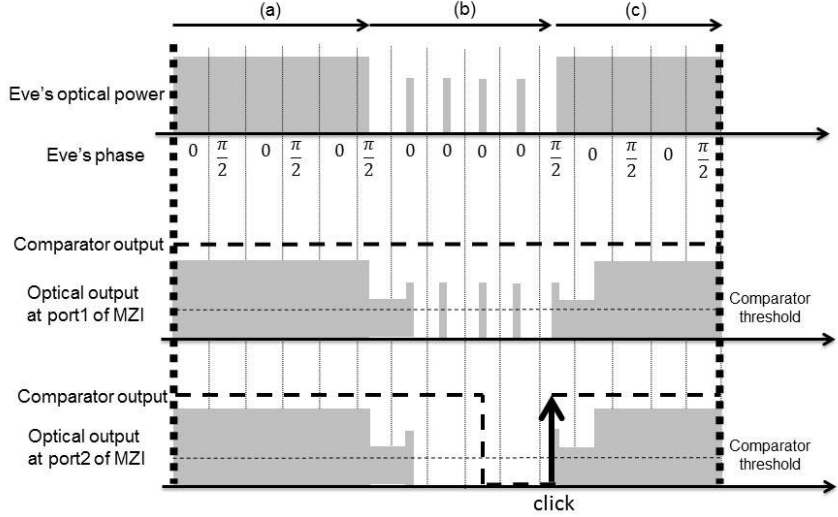

Fig. 3. Pulse pattern in tailored bright illumination attack against DPS-QKD [6]. (a) Blinding phase where both detectors get blind. (b) Recovery phase where one of two detectors connected to port1 becomes clickable, and the other stays blind. (c) Re-blinding phase where detector connected to port1 gets blind again after click, the other stays blind.

To defeat this kind of attack, we propose a countermeasure by slightly modifying the setup of DPS-QKD. As shown in Fig. 5, we connect two SSPDs at the each output port of the Mach-Zehnder interferometer with a 50:50 coupler, and monitor the rate of coincidence detection at the each of the pairing detectors. Bob monitors and estimates the conditional coincidence probability that one of two detectors in an output port clicks given the click of the other detector in the same output port. The conditional coincidence rate ($CCR$) is estimated by

$$CCR = \frac{1}{4}\mu T\eta + d, \qquad (1)$$

where $\mu$ is the average number of photon per pulse, $T$ is the transmittance, $\eta$ is the quantum efficiency of the single photon detector, and $d$ is the dark count probability. Note that $CCR$ is linear of $\mu$, $T$ and $\eta$ because we concern the coincidence rate conditioned on a click of one of the two detectors. In the normal operation, the coincidence detection is pretty rare. However, in the tailored bright illumination attack Eve necessarily inputs the bright pulse at the time slot at which she wants to make the detector click and change it into the blinding state again. It follows that the probability of the simultaneous detection at the pair of two detectors should significantly increase compared to the normal condition. When Eve attacks all the bits, $CCR$ should be close to 1, and then Bob can detect this attack. By discarding all the bits, Bob can defeat this kind of attack. When Eve only attacks a part of the bits, Bob should be able to notice the change of the coincidence rate. By comparing the estimated conditional coincidence rate, $CCR_{est}$, and the obtained conditional coincidence rate, $CCR_{exp}$, Bob can estimate how much bits are attacked by Eve. In the privacy amplification process, Alice and Bob assume all information of the attacked bits is leaked to Eve. Based on the general individual attack against DPS-QKD [16], Alice and Bob can estimate the final key length, $K_{sec}$, according to

$$K_{sec} = K_{sift}[1-(CCR_{exp}-CCR_{est})-[(1-(2\mu(1-\eta T))\log_2(1-e^2-\frac{(1-6e)^2}{2})] \quad (2)$$
$$-f(e)\{-e\log_2 e-(1-e)\log_2(1-e)\}],$$

where $K_{sift}$ is the sifted key length, $\mu$ is the average number of photon per pulse, $T$ is the transmittance, $\eta$ is the quantum efficiency of the single photon detector, $e$ is the quantum bit error rate, and $f(e)$ characterizes the performance of the error correction algorithm.

Note that in the real situation, we may have to take into account the experimental imperfection such as the fluctuation of the conditional coincidence rate, the effect of the finite length of the sifted key, inequality of the quantum efficacy of the single photon detectors. Further studies are needed to take into account these imperfections.

In this countermeasure described above, we assume 50:50 coupler split the blinding light equally. However, it may possible for Eve to blind one of two detectors in a pair by changing the wavelength of the blinding light. In such a case, we can defeat this kind of attack by putting the optical band bass filter in front of the Mach-Zehnder interferometer. Especially, in our field experiment we have already set an optical band bass filter to suppress stray light from other optical fibers [18]. Another way is to check the detection statistics of all four detectors.

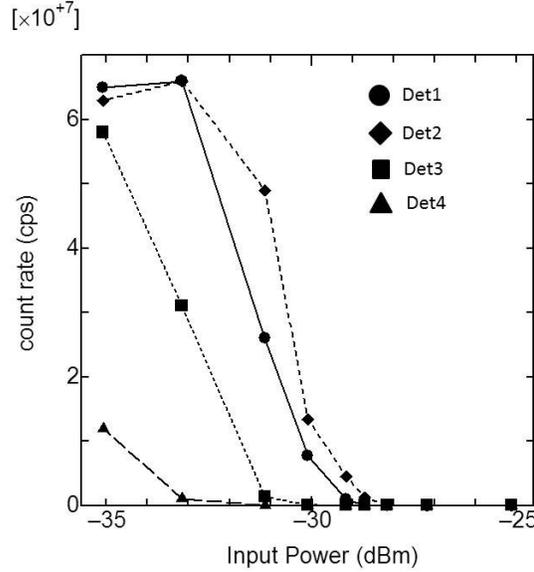

Fig. 4. Count rate as a function of the power of the input light.

## 4. Experimental Setup

First of all, we checked the characteristics of our SSPDs when a bright light was injected. To understand the characteristics in the following QKD experiment, we injected a 1551nm pulse stream with a 1 GHz repetition rate instead of CW light. We measured the count rate for various input light powers, and the experimental results are shown in Fig. 4. Although the characteristics of the four detectors in high count rate regime were slightly different from each other due to their individual difference in kinetics inductance, we confirmed that the minimum power to completely blind all the SSPDs was ~-25dBm light (~$2.5\times10^4$ photons/pulse) [19].

This power is 1000 times lower than the original demonstration, thanks to the reduction of wasting light by using a pulse stream for blinding light and high quantum efficiency (~10%) of our SSPDs [20].

Next we have experimentally confirmed the feasibility of our countermeasure. Figure 5 shows our experimental setup [21]. First, we explain the setup of normal operation shown in Fig. 5(a). In Alice's site, a 1551 nm continuous light wave from a semiconductor laser is converted into a 70 ps width pulse stream with a 1 GHz repetition rate using a $LiNbO_3$ intensity modulator. Each pulse is randomly phase-modulated by $\{0,\pi\}$ with a $LiNbO_3$ phase modulator driven by the random bit signal from the FPGA board [21]. (In the following attack emulation, 0-π static phase modulation was used instead of random phase modulation.) The optical pulse is attenuated down to 0.2 photons/pulse. Then, 18dB attenuation is applied to emulate the fiber transmission. Then, the optical pulse is input into to Bob's receiver.

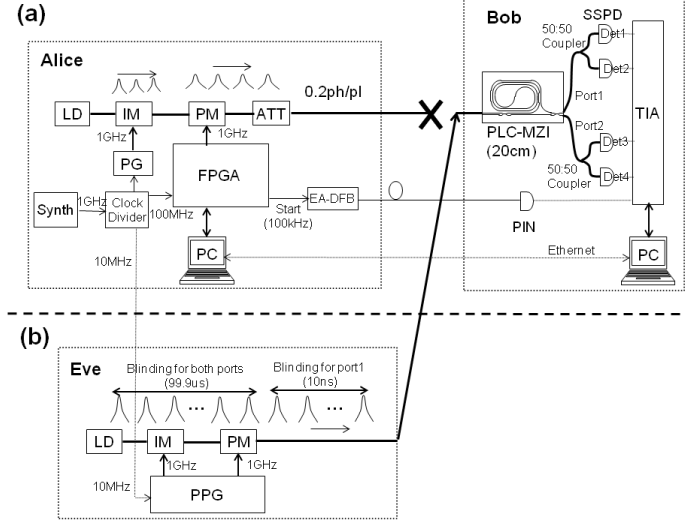

Fig. 5. Experimental setup.

In Bob's site, the 1 GHz pulse stream is input to a PLC-MZI. Two pairs of SSPDs [20] (Det1 and Det2, Det3 and Det4) are connected to each of the output ports of Mach-Zehnder interferometer with a 50:50 coupler. The detected signals are input into a time interval analyzer (TIA) via a logic gate to record the photon detection events. Bob's PC retrieves the data from TIA and sends the time information to Alice following the DPS-QKD protocol. Then, Alice and Bob generate sifted keys.

Figure 5(b) shows the emulation setup of the tailored bright illumination attack. In Eve's site, a 1551 nm continuous light wave from a semiconductor laser is converted into a 300 ps width pulse stream with a 1 GHz repetition rate using a $LiNbO_3$ intensity modulator. Each pulse is appropriately phase-modulated by $\{0,\pi\}$ with a $LiNbO_3$ phase modulator driven by the pulse pattern generator (PPG). To synchronize PPG with Alice's setup, 10MHz clock is provided from Alice's setup. For the blinding lights (pulses), the phase modulation pattern of $\{0,0,\pi,\pi\}$ is repeatedly applied to the trains of pulses such that half of the pulses go to the one of the two output ports of Mach-Zehnder interferometer. By applying the above phase modulation pattern, we need not to treat π/2 phase-modulations. Note that we use the pulse stream instead of originally proposed large duration pulse due to the following reasons. One is to decrease the power of the blinding pulse. The other is to sharply apply the phase modulation to control the target detectors. After the 9.99 μs blinding pulse stream, 10 pulses (10ns) with no phase modulation were input into Mach-Zehnder interferometer. During this 10ns period, all pulses go to the ouputport of the port1 (Det1 and Det2) of Mach-Zehnder interferometer, and the SSPDs connected to port1 stay blind. The SSPDs connected to port2 (Det3 and Det4) become clickable. Thus, Eve can deterministically fire Det3 and Det4 at the beginning of the next blinding pulse stream.

With this setup we estimated the quantum bit error rate (QBER) and the rate of coincidence detection of Det3 and Det4. The QBER was 0.0%, which means Eve can perfectly control the Bob's detection, and the conditional coincidence rate was 0.996. The count rate of Det1, Det2, Det3 and Det4 were 0, 0, 99.9k, 99.9k cps respectively. The average power of blinding light at the input of SSPDs was −25.85 dBm on average. We also measured the QBER and the coincidence rate under the normal operation. QBER was 3.2% , which mainly comes from the dark count of detectors and the imperfection of the Mach-Zehnder interferometer, and the conditional coincidence rate of Det1-Det2 and Det3-Det4 were $1.4 \times 10^{-4}$ and $5.4 \times 10^{-5}$ respectively.

The estimated conditional coincidence rate based on Eq. (1) was $5.0 \times 10^{-5}$, which is almost reasonable compared to the experimental results. Thus, we have confirmed that the tailored bright illumination attack is possible, and our countermeasure works well to detect this kind of attack by monitoring the coincidence rate.

## 6. Conclusion

In summary, we propose a countermeasure against tailored bright illumination attack for DPS-QKD. Alice and Bob can defeat this kind of attack by monitoring a rate of the coincidence detection at a pair of SSPDs. We also experimentally confirmed the threat of the so-called tailored bright illumination attack proposed by L. Lydersen et al. and demonstrated the feasibility of our proposed countermeasure. In this paper, we concentrate on our DPS-QKD protocol. However, this kind of countermeasure must work on the other QKD system employing the other protocols.